\documentclass[aps,prc,groupedaddress,twocolumn,showpacs,amsmath,amssymb]{revtex4}
\usepackage{graphics}
\usepackage{dcolumn}
\usepackage{longtable}

\begin{document}

\title{Analysis of alpha-induced reactions on $^{151}$Eu below the Coulomb barrier}

\author{V.~Avrigeanu}
\email{vavrig@ifin.nipne.ro}
\author{M.~Avrigeanu}
\affiliation{ ``Horia Hulubei'' National Institute for Physics and Nuclear Engineering, P.O. Box MG-6, 077125 Bucharest-Magurele, Romania}

\begin{abstract}
Novel measurements of $(\alpha,\gamma)$ and $(\alpha$,n) reaction cross sections on the target nucleus $^{151}$Eu, close to the reaction thresholds, support the setting up of recent parameters of the $\alpha$-particle optical model potential below the Coulomb barrier. A better understanding of the $\alpha$-particle optical potential at these energies leads to a statistical model analysis of additional partial cross sections that were formerly measured but not considered within the model analysis. On this basis we have tentatively assigned a modified $J^{\pi}$=9$^-$ spin and parity to the 22.7-h isomer in $^{154}$Tb.
\end{abstract}


\maketitle

The cross sections of the $^{151}$Eu$(\alpha,\gamma)$$^{155}$Tb and $^{151}$Eu$(\alpha$,n)$^{154}$Tb reactions have been recently measured at energies relevant for the astrophysical $\gamma$-process, namely between 12 and 17 MeV, in order to extend the related experimental database towards the heavier mass region \cite{gg2010}. These results were compared with predictions of statistical model calculations and it was found that the calculations using the well-known optical potential by McFadden and Satchler \cite{lmf66} overestimate the cross sections by about a factor of 2. A careful sensitivity analysis performed at the same time has shown that this discrepancy is caused by the inadequate description of the $\alpha$+nucleus channel which is regarded as a current challenge in nuclear astrophysics \cite{gg2010}.  

In the meantime an analysis of all available $\alpha$-particle induced reaction cross sections on nuclei within the mass number range 45$\leq$$A$$\leq$197, below the Coulomb barrier, has been carried out \cite{ma2010} leading to an optical model potential (OMP) which describes the $\alpha$-particle elastic scattering at low energies as well. Taking advantage of both elastic-scattering and $\alpha$-particle induced reaction data systematic analysis, the energy dependence of the surface imaginary potential depth has been proved to be essential for the understanding of the $\alpha$-particle interaction behavior below the Coulomb barrier. This OMP (Table I of Ref. \cite{ma2010}) has been used in the present work within statistical model (SM) calculations of the $^{151}$Eu$(\alpha,\gamma)$$^{155}$Tb and $^{151}$Eu$(\alpha$,n)$^{154}$Tb reaction cross sections. We use the same approach \cite{ma2010} and consistent set of local SM parameters that have been established or validated on the basis of independent experimental information for, e.g., neutron total cross sections, $\gamma$-ray strength functions based on neutron-capture data, and low-lying levels and resonance data. The results of this calculation are found in close agreement (Fig. 1) with the measured $(\alpha$,n) reaction cross sections \cite{gg2010}. A slightly different case is that of the $(\alpha,\gamma)$ reaction, its calculated cross sections being also monitored by the $\gamma$-ray strength functions. Nevertheless, the use of the related systematic model parameters (Sec. II.B of Ref. \cite{ma2010}), corresponding finally to a $s$-wave neutron resonances radiative width $\Gamma_{\gamma 0}$$\sim$150 meV of the compound nucleus $^{155}$Tb, makes possible the accurate description of also the $^{151}$Eu$(\alpha,\gamma)$$^{155}$Tb reaction cross sections.  

\begin{figure} [b]
\resizebox{0.95\columnwidth}{!}{\includegraphics{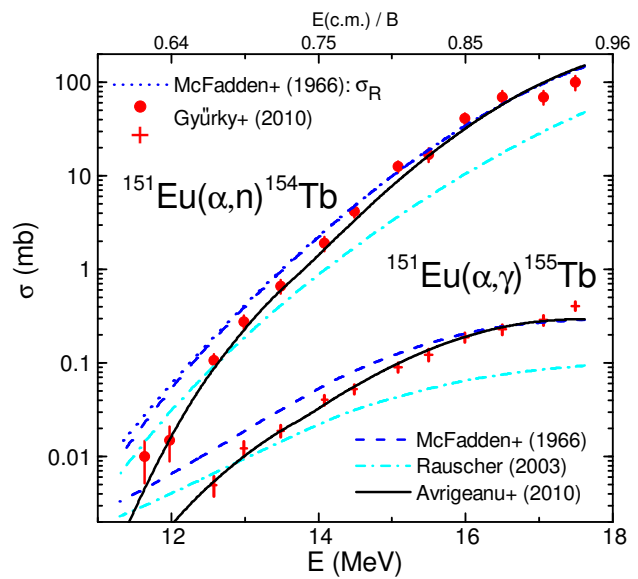}}
\caption{\label{Fig1}(Color online) Comparison of the measured \cite{gg2010} and calculated $(\alpha,\gamma)$ and $(\alpha,n)$ reaction cross sections for the target nucleus $^{151}$Eu, using the OMPs of Refs. \cite{lmf66} (broken curves), \cite{tr03} (chain), and Table I of Ref. \cite{ma2010} (full). The calculated $\alpha$-particle total reaction cross sections corresponding to the OMP of Ref. \cite{lmf66} are shown as well (dotted).}
\end{figure} 

On the other hand, Fig. 1 also shows the SM calculated cross sections corresponding to the OMPs of McFadden and Satchler \cite{lmf66} and Rauscher \cite{tr03}, which are discussed in Ref. \cite{gg2010}, while the remaining  model parameters are unchanged. These potentials obviously fail to describe at the same time both $(\alpha,\gamma)$ and $(\alpha$,n) reactions on $^{151}$Eu, even when some reduction or enhancement factors are used. At the same time the total reaction cross sections corresponding to one \cite{lmf66} of these OMPs are also shown in Fig. 1 in order to point out the key role of the $\alpha$-particle OMP for the model predictions of the dominant $(\alpha$,n) reaction cross sections. 
  

Once the understanding of the $\alpha$-particle OMP is improved, the recently measured partial $(\alpha$,n) reaction cross sections to the ground state (g.s.) and two long-lived isomeric states of the residual nucleus $^{154}$Tb \cite{gg2010} can also be used in this analysis. A SM calculation of these cross sections would be rather questionable because of several peculiar issues that characterize this odd-odd neutron-deficit isotope, the only stable isotope of Terbium being $^{159}$Tb. Firstly, only excitation energy limits of the two isomers are known, namely $\le$25 keV \cite{cwr09} and (200$\pm$150) keV \cite{ga03} for the m1 and m2 isomers, respectively.  Secondly, there is no data concerning their feeding by $\gamma$-ray cascades from the discrete levels above them, which are populated at their turn within the $(\alpha$,n) reaction through side feeding and decay of the continuum of excited states. Actually, beyond the two isomers, only several high-spin level bands are known, excited through heavy-ion induced reactions of the type (HI,x5n) \cite{cwr09}. Consequently, we first started with a rough calculation of the above-mentioned partial cross sections using only the three discrete levels of the residual nucleus $^{154}$Tb for which these data had been measured (Table 3 of Ref. \cite{gg2010}). The continuum of excited states above the three discrete levels has been described by the back-shifted Fermi gas (BSFG) nuclear-level density parameters $a$=18.15 MeV$^{-1}$ and $\Delta$=-0.95 MeV. These parameter values correspond to the BSFG parameter systematics \cite{va02} and fit of the low-lying discrete levels \cite{ensdf} of the similar isotopes $^{156,158}$Tb. We have taken advantage of $N_d$$\approx$18 discrete levels of $^{156}$Tb, up to the excitation energy $E_d$$\approx$405 keV, which are also rather close to the level scheme of $^{158}$Tb better known up to even higher energies. The calculated partial $(\alpha$,n) reaction cross sections are shown in Fig. 2(a). The agreement obtained for the dominant m1 isomer has been expected due to the similar one found for the $(\alpha$,n) reaction total cross section (Fig. 1). On the other hand, the calculated cross sections for the g.s. and m2 states are different by factors from 2 to 10 with respect to experimental data, while the order of their excitation functions is reversed.

\begin{figure} [b]
\resizebox{0.82\columnwidth}{!}{\includegraphics{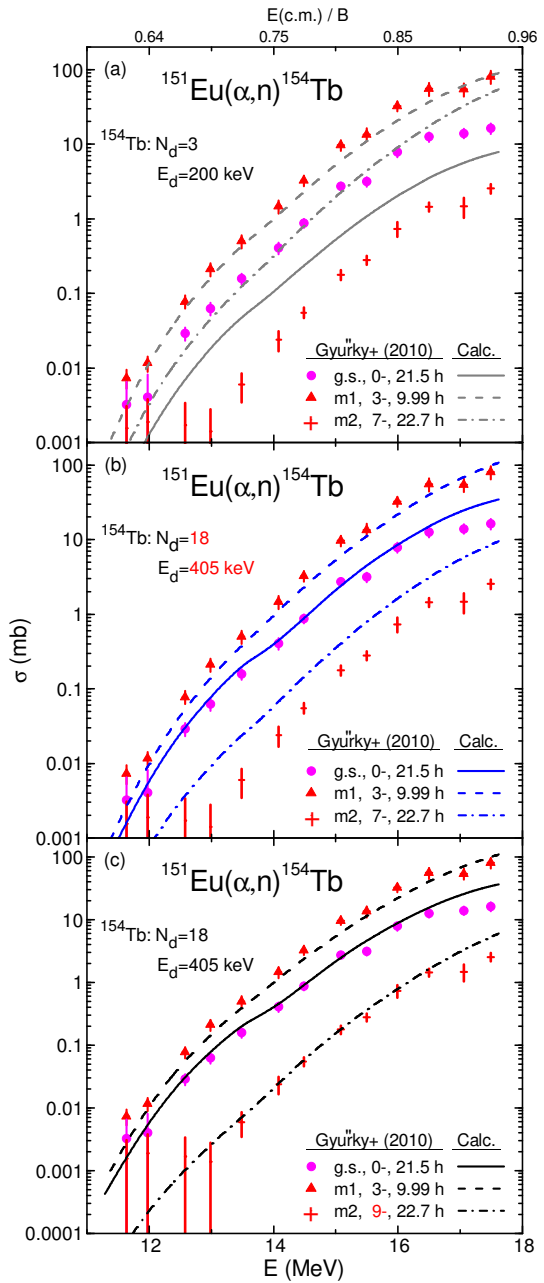}}
\caption{\label{Fig2}(Color online) Comparison of the measured \cite{gg2010} and calculated partial $(\alpha$,n) reaction cross sections to the ground state (g.s.) and m1 and m2 isomeric states of the residual nucleus $^{154}$Tb using discrete level schemes of this nucleus including (a) only the three long-lived levels \cite{cwr09}, (b) a number of $N_d$=18 levels up to the excitation energy $E_d$=405 keV, on the basis of the $^{156}$Tb nucleus structure \cite{cwr03}, and (c) an additional assumption of 9$^-$ m2 isomer \cite{gg2010}.}
\end{figure}

As we aimed to use a more realistic low-lying discrete level scheme of the residual nucleus $^{154}$Tb in the SM calculation, the following  assessment has imposed itself as the only viable solution. Thus, the $N_d$=18 low-lying levels of $^{156}$Tb, up to the excitation energy $E_d$=405 keV, is used within the present SM calculation as the discrete level scheme of the residual nucleus $^{154}$Tb. The only change done in this respect is the replacement of the long-lived 3$^-$, (7$^-$) and (0$^+$) levels of the $^{156}$Tb nucleus \cite{cwr03} by the g.s. and m1 and m2 isomers of $^{154}$Tb. An additional assumption concerns the g.s. parity, a negative value being used in agreement with the structure of the other odd-odd Tb isotopes \cite{ensdf}. Moreover, the spin, parity, and decay of the $^{156}$Tb levels just above the $N_d$=18 have been considered for the few lower levels of $^{156}$Tb nucleus without the corresponding assignments \cite{cwr03}. The results displayed in Fig. 2(b) show a marked improvement for the g.s. population while the assumption of the g.s. negative parity plays no role in this respect. However, the partial cross sections of the m2 isomer are still overestimated by an average factor of $\sim$2. Furthermore, this difference has been found to be rather insensitive to any change of other SM parameters as the nuclear-level density or $\gamma$-ray strength functions. 

Under these circumstances a final remark concerns the second isomer spin. A former suggestion of a 9$^-$ m2 isomer shown in Fig. 1 of Ref. \cite{gg2010} corresponds indeed to reduced partial cross sections for its population. This assumption is confirmed by a final SM calculation of our work, the only additional change being the replacement of the 7$^-$ m2 isomer by a 9$^-$ one. The results shown in Fig. 2(c) describe well the measured data except for the two data points at the highest energies. The reason of this sudden decrease of the g.s. and m2 isomer partial cross sections is less clear, no new reaction channel being just opened at these energies while the $\alpha$-particle  total reaction cross section is continuously increasing (Fig. 1). Nevertheless, taking into account the last assumption of a 9$^-$ m2 isomer, one may consider that the new $(\alpha,\gamma)$ and $(\alpha$,n) reaction cross section measurements on the target nucleus $^{151}$Eu \cite{gg2010}, close to the reaction thresholds, support the setting up of the $\alpha$-particle OMP \cite{ma2010}.

\bigskip
This work was partly supported by the CNCSIS project PNII-IDEI-43/2008.

\end{document}